\documentclass{IEEEcsmag}

\usepackage[colorlinks,urlcolor=blue,linkcolor=blue,citecolor=blue]{hyperref}

\usepackage[style=ieee,doi=false,isbn=false,url=false]{biblatex}
\usepackage{subcaption}
\usepackage{float}
\usepackage{upmath}
\graphicspath{{figures/}{./}}

\jvol{XX}
\jnum{XX}
\paper{8}
\jmonth{May/June}
\jname{Computer Graphics \& Applications}
\pubyear{2019}

\begin{document}

\sptitle{Department: Head}
\editor{Editor: Name, xxxxemail}

\title{What we talk about when we talk about data physicality}

\author{Dietmar Offenhuber}
\affil{Northeastern University, e-mail: d.offenhuber@northeastern.edu}

\markboth{Department Head}{Paper title}

\begin{abstract}
Data physicalizations “map data to physical form,” yet many canonical examples are not based on external data sets. To address this contradiction, I argue that the practice of physicalization forces us to rethink traditional notions of data. This paper proposes a conceptual framework to examine how physicalizations relate to data. This paper develops a two-dimensional conceptual space for comparing different perspectives on data used in physicalization, drawing from design theory and critical data studies literature. One axis distinguishes between epistemological and ontological perspectives, focusing on the relationship between data and the mind. The second axis distinguishes how data relate to the world, differentiating between representational and relational perspectives. To clarify the aesthetic and conceptual implications of these different perspectives, the paper discusses examples of data physicalization for each quadrant of the continuous space. It further uses the framework to examine the explicit and implicit assumptions about data in physicalization literature. As a theoretical paper, it encourages practitioners to think about how data relate to the manifestations and the phenomena they try to capture. It invites exploration of the relationship between data and the world as a generative source of creative tension.
\end{abstract}


\maketitle

\begin{figure*}[ht]
\centering \includegraphics[width=\textwidth]{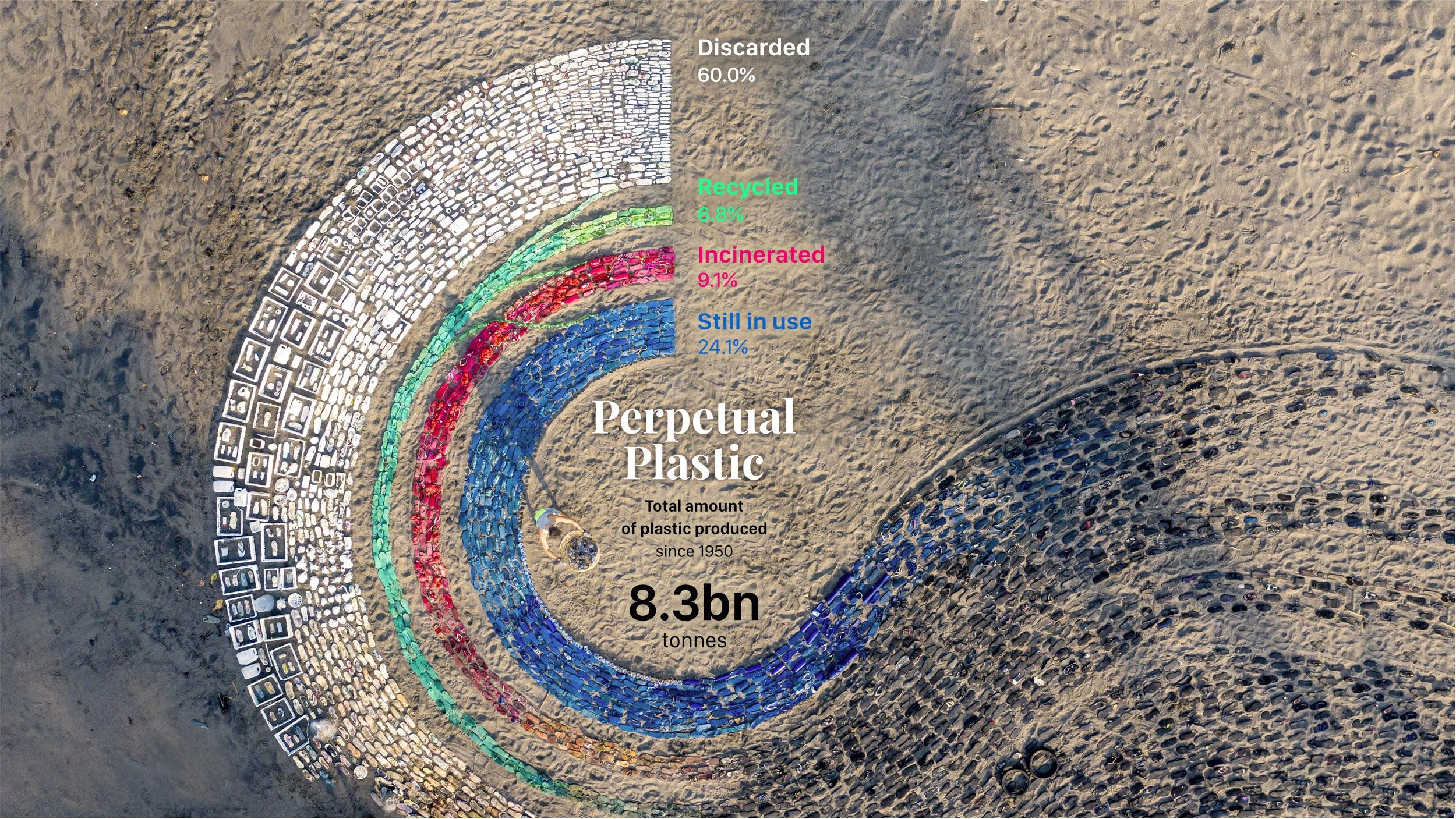}
\caption{Perpetual Plastic by Liina Klauss, Skye Morét, and Moritz Stefaner. The colored bands represent the quantities of plastic by waste stream (disposal, incineration, recycling, in use) and their connections. Source: \url{http://perpetual-plastic.net}}\label{pplastic}
\end{figure*}

\chapterinitial{Introduction}
For most practitioners of information visualization, the concept of data seems unproblematic and trivial: what can a data set be, other than a file with columns and values, keys and attributes? In this paper, I argue that data physicalization forces us to expand this narrow understanding of data and their relationship to the world. I hope to show, in theoretical arguments and examples, that the practice of data physicalization transcends the traditional epistemological and representational models of data and requires us to consider ontological and relational perspectives. In that sense, I address a question similar to the one posed by Paul Dourish in his seminal paper “What we talk about when we talk about context,” where he challenges conventional interpretations of \textit{context} as a representational issue~\cite{dourish2004}.

\begin{figure}[ht]
\centering \includegraphics[width=\linewidth]{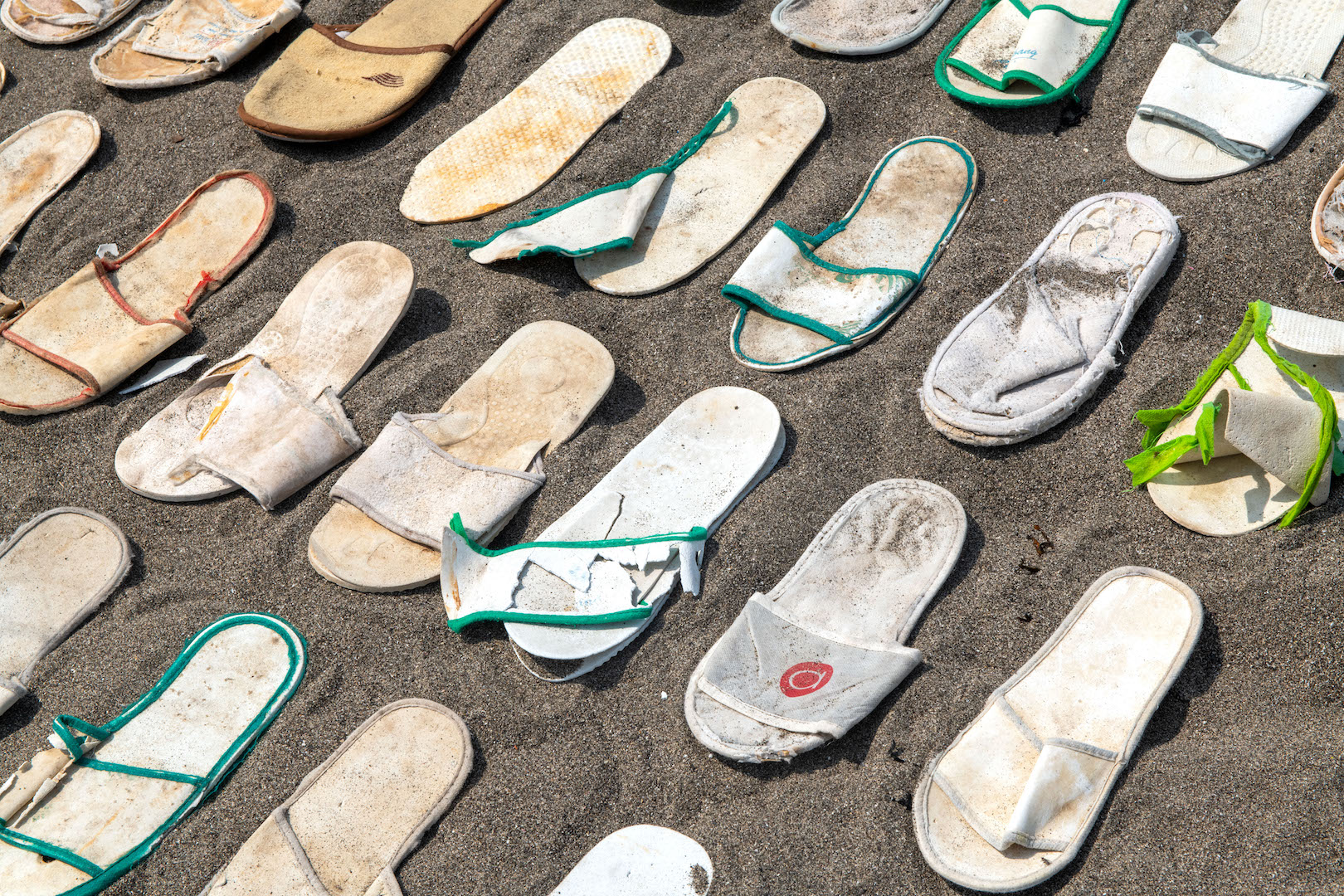}
\caption{Perpetual Plastic (detail). Source: \url{http://perpetual-plastic.net}}\label{plastic}
\end{figure}

My argument can be illustrated with the recent data physicalization project “Perpetual Plastic” by Liina Klauss, Skye Morét, and Moritz Stefaner (Fig.~\ref{pplastic}). The physicalization is situated on a beach in Bali; it consists of plastic debris collected from the beach and arranged into a Sankey diagram. The diagram represents the fate of plastic waste: how much of the plastic has been discarded, incinerated, recycled, and reused. From the conventional logic of visualization, it is not a very efficient display: it uses a lot of material and space to express a handful of numbers, the ratios of the waste stream. But applying such a metric would lead to a complete misreading of the project, even if we ignore its compelling aesthetic and artistic qualities. The physicalization does not just represent data; it \textit{is} data. It consists of the things it diagrammatically represents, plastic marine debris collected during a recent cleanup event from the area around the beach (Fig.~\ref{plastic}). 

Viewed as a nearly complete sample of material data, one could read the installation in any number of ways. One could, for example, examine the different stages of material decay of plastic slippers in the artwork as a result of exposure to the elements, contemplate the unsettling durability of nylon nets, estimate the impact of tourists from the number of sunscreen bottles, and so on. The installation establishes a wide range of meaningful relationships with its surroundings: with the beach from where the plastic was collected, wind and water disturbing and eventually destroying the arrangement, and the recipients, who might awkwardly hold half-empty water bottles as they watch a video of the installation in an exhibition. It is fair to assume that many viewers will not correctly decode the Sankey diagram in its physical form; it works mostly as an icon. The five numbers mapped to ribbons with different widths and colors are a container that holds the more relevant data of the project: the material specimen and the contextual relationships established in the act of making the physicalization.

Is it justifiable to describe the assemblage of items that make up the installation as data? They are arguably more than decorative elements. They are also more than clarifying illustrations since they convey original information about the phenomenon; the installation does not prescribe that it is to be read solely as a Sankey diagram. Similar to a traditional data collection process, the items have been collected and categorized by shape and color. Whether they have been recorded and digitized before their assembly on the beach is largely irrelevant for the physicalization.

However, the physical data found in \textit{Perpetual Plastic} differ from digital data in several ways. Digital data are discrete and atomistic. The total amount of information can be quantified. Physical data, on the other hand, are not abstracted and contain unquantifiable amounts of information. Any number of digital data sets could be generated from the sample, depending on which of the items' characteristics are of interest. The missing step of abstraction makes physical data contextual and full of potential meaning. An installation that relies on physical data has to establish a context for their interpretation.

\begin{figure}
\includegraphics[width=\linewidth]{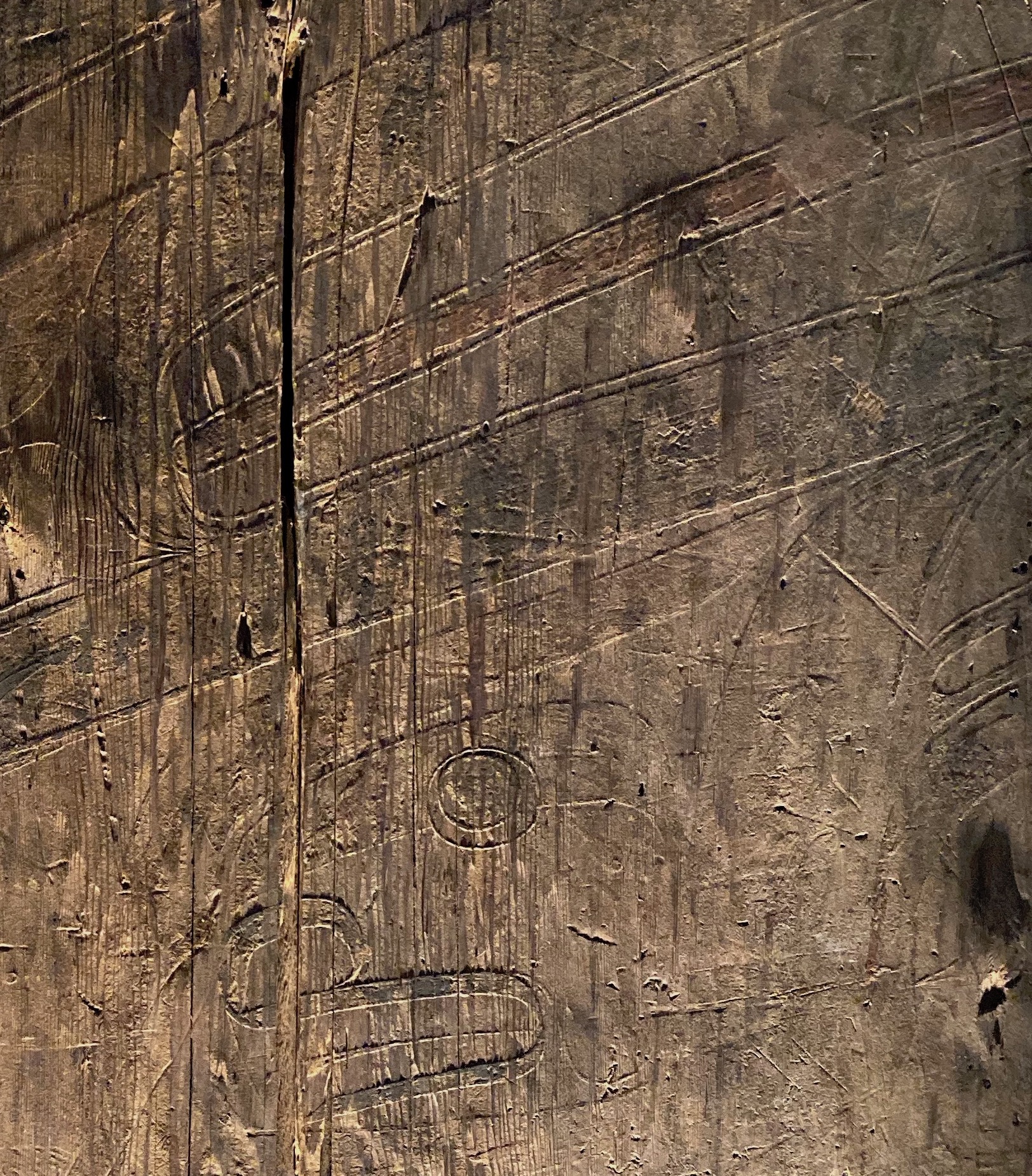}
\caption{Subtle indigenous wood carvings, invisible under normal light, made visible by oblique lighting. Museum of Anthropology, Vancouver. Photo by the author}\label{carving}
\end{figure}

A similar issue arises for physical expressions of traditional data sets. Data values encoded into physical materials come with additional properties such as weight, color, texture, and smell. Conversely, small data differences may disappear in the material's surface texture and the traces of fabrication. Many aspects are outside the designer's control: given the right environmental lighting, small variations may appear exaggerated; small bumps turn into mountains (Fig.~\ref{carving}).

Isolated by the margin of the page or the edge of the screen, traditional visualizations are less dependent on their surroundings. Their visual language is designed to be \textit{monosemic} (in the words of Jaques Bertin) and universal rather than contextual. Visualization owes its central concept of \textit{retinal variables} to cartographic communication theory, which frames charts and maps as a formal system of communication~\cite{bertin1983}.  This framing, however, is not uncontroversial. The critical cartography movement raised doubts about whether maps \textit{can} and \textit{should} fulfill the expectation of a formally consistent communication system, arguing that while maps are practically never consistent, they convey much more information through context than through the codes specified in their legend~\cite{crampton2005}. Similar critiques apply to data visualization, and the \textit{polysemic} qualities of physicalizations bring them into the focus of attention.

Data physicalization brings data from the unambiguous symbolic space into the real world, where data is a more complicated affair. As the physical manifestation of a data set becomes more elaborate and sensorily rich, data and display cannot be neatly separated. This is not only due to the lack of control of the designer but also due to the lack of visual conventions familiar to the recipient. In the case of data visualizations, the visual conventions of data literacy have disciplined our views so that we ideally are able to abstract the message of a chart from its physical properties. As data physicalization does not (yet) have such conventions, every encounter with a data object becomes a critical inquiry into materiality, context, and the process of making. I would argue for embracing this unfamiliarity and exploring its potential for critical reflection.

\section{Approach}

The goal of this paper is to clarify theoretical questions that are foundational to data physicalization, but have not been sufficiently addressed yet. To this effect, the paper proposes a simple schema to examine the relationship between physical visualizations and the various forms of data that shape their reception. To investigate how data physicalizations embody data, I will address two questions:

\begin{enumerate}
\item Which perspectives on data does the physicalization literature invoke?
\item Which perspectives on data does the practice of data physicalization mobilize?
\end{enumerate}

To answer the first question, I analyze the arguments in the literature about data physicalization and related fields such as ubiquitous computing, tangible and ambient media. Since HCI literature has only recently started to focus on data and the distinction from the concept of information are often blurry and implicit, I used methods of discourse analysis to examine the arguments in these papers in their broader context. Paying close attention to wording, I coded and categorized relevant statements and compared them with theoretical frameworks from the fields of philosophy, science and technology studies, the digital humanities, and the history of science. 

Addressing the second question, I propose a two-dimensional conceptual space that compares different physicalization projects with respect to their perspectives on data. One axis of this space distinguishes between \textit{epistemological} and \textit{ontological} conceptualizations of data, while the other differentiates between \textit{representational} and \textit{relational} perspectives. The epistemological-ontological continuum compares definitions of data as a product of the human mind versus data as observer-independent patterns in the world. The representational-relational axis focuses on how data express meaning. It distinguishes between views of data as representational sign-vehicles and views of data as implicit relationships and material transformations. To clarify the aesthetic and conceptual implications of the different perspectives, the paper compares examples of data physicalization for each quadrant of the continuous design space. To source examples of data physicalizations, I relied on the examples discussed on the data physicalization wiki~\footnote{See:~\url{https://dataphys.org}} and added further examples from media art, design, and history.

\section{Data Abstraction and Embodiment}

\begin{figure}
\includegraphics[width=\linewidth]{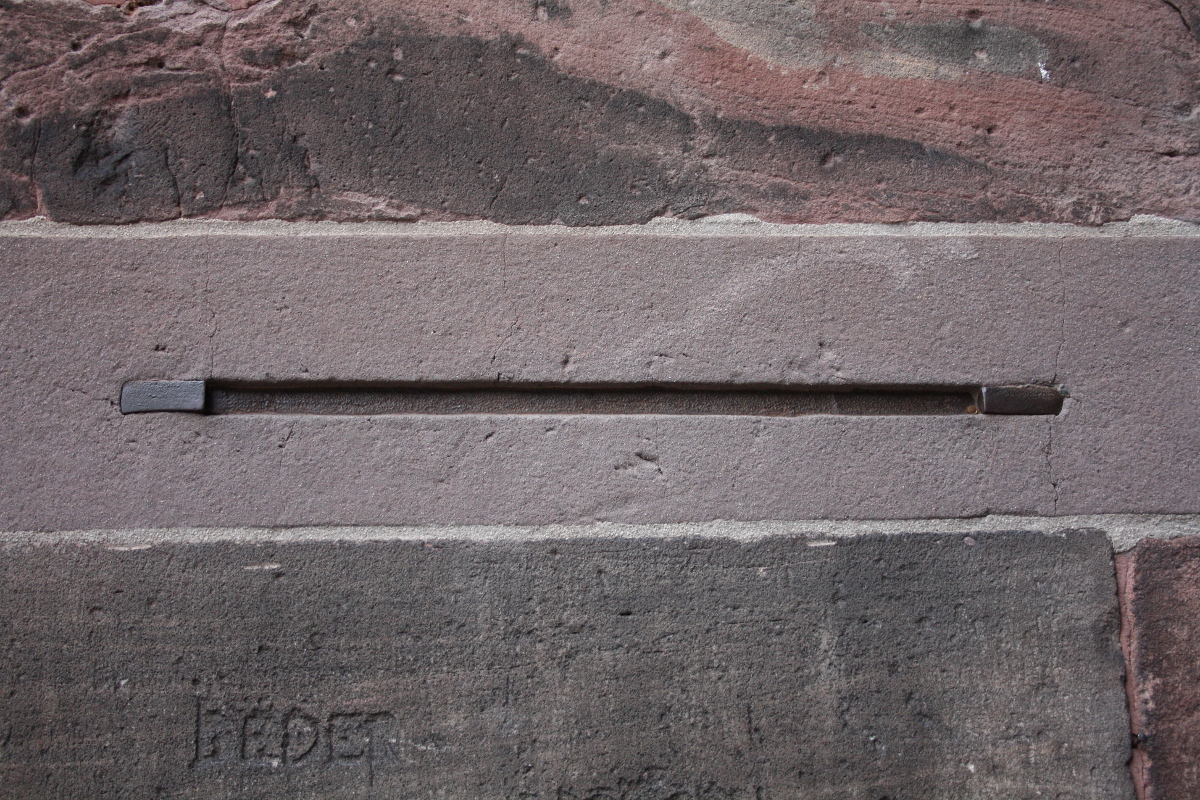}
\caption{Historical measure of length (one cubit) on the wall of the cathedral in Freiburg, Germany. Photo: Birgit Rucker }\label{measure}
\end{figure}

The confusion between data as abstract-platonic entities and physical embodiments runs throughout the history of science and takes different forms in various disciplines. Again, an illustrative example might help. In his study of measurement in the middle ages, historian Emanuele Lugli discusses historical measures attached to or carved into the gates, cathedrals, or squares of medieval cities where they served as the basis for measurement in commerce and architecture~\cite{lugli2019}~(Fig.~\ref{measure}). As tangible units for comparison, the measures played a central role in creating and verifying data records in land sales, commerce, and architecture. The lengths of the measures themselves, however, were subject to considerable uncertainty and regional differences. The standard of the \textit{braccio} varied widely across Italian cities and fluctuated in time due to trade, physical degradation or manipulation, and political decisions to redefine or introduce new measures. To maintain trust in the measures, according to Lugli, required authorities to establish the notion of the measure as something abstract --- embodied in the measure on display, but not identical with it. He points out that the public displays of measures found in the walls of palazzi in Bologna or Padua are executed as incisions rather than attached objects. Their negative space emphasizes the abstract nature of the measure~\cite[p. 68]{lugli2019}. Political power always manifests itself in the ability to make things disappear --- by making them seem natural and uncontroversial. The naturalization of measures has been so successful that we tend to lose sight of their material embodiments, which are ultimately those that `count.' Measuring remains a material practice involving physical tools and bodily labor, subjective judgment, and the repeated confrontation of discrepancies.

It is not a large leap to compare this process to the interpretation of a physicalized data set. The designer, familiar with the underlying data, sees the physicalization as a straightforward representation. The situation is different for the observer --- the object embodies an absent and unfamiliar data set in the same way the incision in stone embodies the abstract measure. Like the measure, the physicalization can become a data embodiment subject to various actions --- it may be passed around, felt with fingertips, weighted, and so on. It may not be clear whether a crease in its surface is a feature of the data set or a trace of the fabrication tool. Yet, the object's material presence, including all its qualities and contextual relationships, becomes identified with the data set.

To address the different roles of data in physicalization, it is necessary to review a few theoretical foundations. Instead of relying on a single authoritative definition of data, I will compare perspectives that highlight different aspects of the concept. Some properties are shared by all of these perspectives. Compared to the more metaphysical concept of information, data have a tangible presence as concrete (arti)facts. Data become useful through their ability to move between contexts. They can be passed around between people, transformed, mapped, and interpreted beyond the purpose for which they were collected. They are assumed to have some relationship to the world; describe a phenomenon, truthfully or not. Data are not considered inherently meaningful, which also differentiates them from the concept of information. Luciano Floridi has defined data as a single difference, a lack of uniformity~\cite{floridi2011}. His general definition of information (GDI) is built on the concept of data: “information equals data plus meaning.” But in order to be useful in different practices, this broad definition needs to be concretized and operationalized, addressing different aspects of the relationship between datum, observer, and the world.

\subsection{Epistemological vs. Ontological Perspectives}
Ontology asks the question “what is” while epistemology asks “how can we know?” \textit{Epistemic perspectives} emphasize the role of data as products of the human mind: interpreted observations captured by the senses or a technical apparatus. They stress the roles and limits of human judgment.

\textit{Ontological perspectives}, in contrast, understand data as physical things --- as patterns in the world that exist independently of the human mind. Ontological definitions view data as proto-epistemic entities embodied in DNA or in the order of geological strata. 

Epistemological and ontological perspectives exist on a continuum. We can imagine data generation as a process that starts with the manipulation of matter and ends with the production of epistemic entities. Different points can be chosen along this continuum to mark the moment when data come into being.

\begin{itemize}
\item Physicist Ralf Landauer argues that information is a physical entity. Since quantum physics poses hard limits on what can be computed, information is never independent of its physical manifestation~\cite{landauer1999}.

\item Ian Hacking defines experimental data as the physical marks produced by an instrument, which then serve as the basis for a variety of practices in and outside the laboratory~\cite{hacking1992}.

\item Bruno Latour characterizes data (\textit{inscriptions} in his terminology) as \textit{immutable mobiles} --- aspects of the world that are captured, preserved and made transportable in order to become useful in arguments~\cite{latour1990}.

\item Hans Jörg Rheinberger locates the beginning of data at the point where material traces are translated into records that can be archived and transmitted~\cite{rheinberger2011}. He describes scientific artifacts such as lab samples as \textit{epistemic things} that embody hypotheses, models, and theories; and are therefore epistemically loaded.

\item Moving towards the epistemological side of the spectrum, data in the social sciences typically involve a higher degree of ordering, cleaning, and interpretation. A data set is typically understood as a rectified synthesis of various notes and records (e.g. collected by field researchers).

\item Definitions of data in the humanities tend to be located on the epistemological end of the spectrum. Johanna Drucker emphasizes the many ambiguities and judgment calls involved in manual data collection, characterizing data (she prefers the term \textit{capta}) as epistemic products of human interpretation~\cite{drucker2011}.
\end{itemize}

In summary, the positions differ in how much the concept of data depends on a human observer. Ontological definitions view data as latent and implicit, while epistemological definitions regard data as explicit and textual. The epistemic-ontic axis could therefore be interpreted as the relationship between data and the human mind.

\subsection{Representational vs. Relational Perspectives}

We can further differentiate between representational and relational models of data. While the previous axis was concerned with the amount of interpretation involved in the production of data (data and mind), this axis is concerned with how data express information about phenomena (data and world).

In the \textit{representational model}, data are signs that point to features of the world. A datum is considered an abstract description of some aspect of reality, such as a number expressing the temperature at a certain place and time. The representation is considered context-independent: the datum remains the same, whether it is written on paper or included in a digital spreadsheet. It is not necessarily specified how the datum relates to the phenomenon---meta-data would address that issue. The only relevant question with regard to the datum is whether its relationship to the world is truthful or not.

As scholars have pointed out, the representational model has many shortcomings~\cite{coopmans2014}. The idea of a 1:1 correspondence between datum and a feature of the world is a stretch for many contemporary data practices. It is appropriate for data produced in a physical experiment, but what about document rankings produced by a search engine algorithm? Data can relate to more than one phenomenon or no phenomenon at all (e.g., random numbers). Data fulfill many purposes beyond description --- such as intermediary data produced during simulations or training neural networks. A more fundamental critique considers the semiotic model of the datum as a pointer to be reductive and insufficient. The model is rooted in linguistics and does not account for the material processes and causalities involved in producing data. 

\textit{Relational models} of data address these critiques of representation. In the relational model, a datum is not defined as a reference to the world, but as a set of relationships among material entities. In Sabina Leonelli's definition, anything in the world can be a datum, as long as it is transportable and can be circulated~\cite{leonelli2016}. If the representational model consists of two linked entities, a relational datum can be imagined as a causal chain of material transformations~\cite{latour1999}. This could, for example, include all the steps involved in building a climate model from tree rings, starting with the extraction and preparation of drill cores, to the integration of the measured year rings into a mathematical model. Despite the numerous transformations, the datum remains part of the original phenomenon.

While the representational perspective views data as context-independent and universal, relational data depend on the situation, location, and people involved. The relational perspective accounts for causality and the data collection process. But, as Leonelli points out, also the representational model has practical advantages; it accounts, for example, for the expectation that a data set remains the same when copied or converted into different formats~\cite{leonelli2016}. Its reductiveness and simplifications can be productive for purposes of generalization.

\section{How does the data physicalization community speak about data?}

The young field of data physicalization builds on earlier work in ubiquitous computing, tangible media, and various data art practices and reframes it for the research agenda of information visualization~\cite{jansen2013a, jansen2015}. In this sense, physicalization adopts both the representational model of visualization (mapping data to perceptual variables) and its typical epistemic research questions, such as evaluating the effectiveness of displays for the support of cognitive tasks. However, the framers Jansen and Dragicevic made it clear that they think beyond the limitations of this framework, citing many historical and contemporary examples that fall outside.

Ubiquitous computing, as originally framed by Mark Weiser, popularized relational perspectives on information~\cite{weiser1991}. Discontent with the declarative rigidity of \textit{cyberspace} and its interfaces, Weiser recognized that the experience of \textit{embodied virtuality} involves more than parsing textual information and requires taking implicit cultural practices and contexts into account. Paul Dourish further elaborated the role of \textit{context}, rejecting a representational understanding of the term. He argued that context is not informational, stable, and divisible, but instead relational, dynamic, and emerging from an activity~\cite{dourish2004}.

Hiroshi Ishii's tangible media group continued many themes from early ubiquitous computing literature~\cite{ishii1997, ishii1998}. Initially, the groups' publications promoted a representational model of data that draws literal analogies and correspondences between \textit{bits} and \textit{atoms}. At the same time, however, these early papers expressed the ambition to emancipate material interfaces from being mere instances of the digital world, which they often describe as impoverished. In later work, the group embraced a more explicitly ontological perspective on data. Citing Landauer's dictum that \textit{information is physical}, the group explored interfaces with natural phenomena without involving intermediary steps of digital mediation~\cite{wang2017}. The programmatic idea of~\textit{radical atoms} constitutes a turn to data materiality, considering material qualities as informational agents independent of digital systems~\cite{ishii2012}.

The design literature related to physicalization is often concerned with the modalities of data expression. Vande Moere offered a stark critique of the representational universalism under the “tyranny of the pixel”~\cite{moere2008}. Diagnosing a disconnect between the discreteness of multi-purpose displays and the nature of physical information, he recommends turning towards the “inherent capabilities of many material objects to communicate meaning and functionality by the natural affordances they possess.” Vande Moere also expands beyond the symbolic mapping paradigm by considering a wide range of semiotic relationships between data and display~\cite{moere2009}.

A small but growing literature takes a decisively ontological perspective on data. A popular choice is to use Peircean indexicality as a model for material visualizations that are not based on traditional data but on traces, symptoms, and other natural indicators~\cite{moere2009, offenhuber2015a}. Schofield et al. propose the adoption of indexical strategies that appeared in conceptual art during the 1970s~\cite{schofield2013}. Lockton et al. introduce a design approach that starts with material qualities~\cite{lockton2017}. The model of autographic visualization focuses on the self-inscribing qualities of physical traces and frames the data collection process as a visual practice of revealing such traces~\cite{offenhuber2019d}.

\section{Case Studies}

\begin{figure}[h]
\centering 
\includegraphics[width=1.05\linewidth]{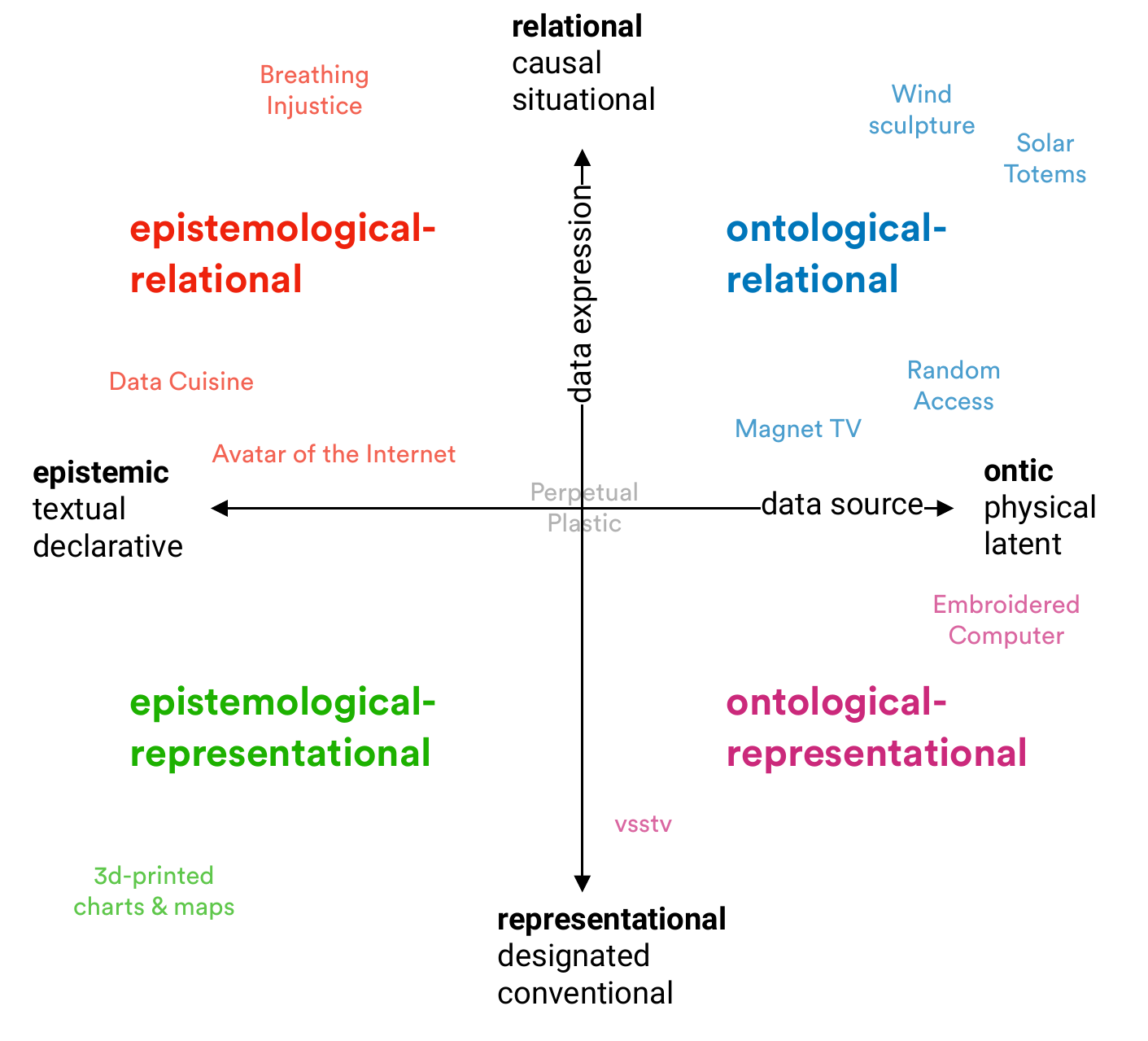}
\caption{Conceptual space spanned by a horizontal epistemic-ontic axis describing the data source and a vertical representational-relational axis describing data expression.}\label{space}
\end{figure}

For the following discussion of physicalization examples, I arrange the discussed perspectives into a two-dimensional, continuous space spanned by the horizontal \textit{epistemic-ontic axis} and the vertical \textit{representational-relational axis}~(Fig.~\ref{space}). The horizontal axis characterizes the data source; the vertical axis the form of data expression. Projects placed on on the left draw from declarative and textual data sources, while the right end of the spectrum draws from analog sources based on latent material qualities and constellations. Projects on the left focus on the epistemic interpretation of a given digital data set, while projects on the right side interrogate the material nature of data. The bottom includes modes of data expression based on explicit data mappings, visual conventions, or metaphors. The top corresponds to expressions that are contextual and situation dependent. The goal of this schema is not to determine a precise location for each project, but to call attention to the nuances and different perspectives on data expressed in these projects. Many works incorporate multiple perspectives. \textit{Perpetual Plastic}, for example, involves a symbolic data set as well as a material data source; includes representational as well as relational aspects.

\subsection{Epistemological / Representational Examples}

\begin{figure}[h]
	\begin{subfigure}{0.525\linewidth}
	\includegraphics[height=3.6cm]{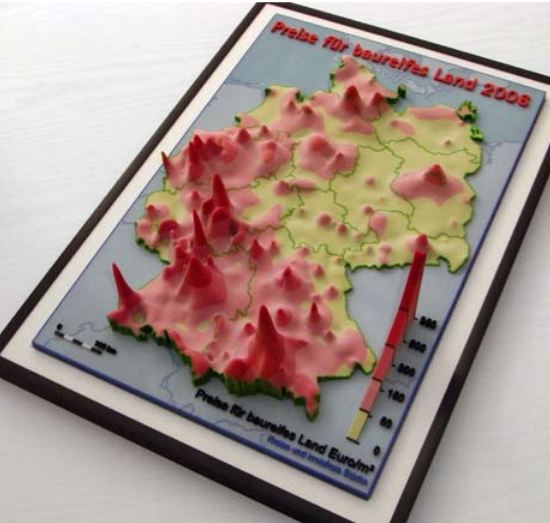}
	\subcaption{}\label{map}
	\end{subfigure}%
	\begin{subfigure}{0.475\linewidth}
	\includegraphics[height=3.6cm]{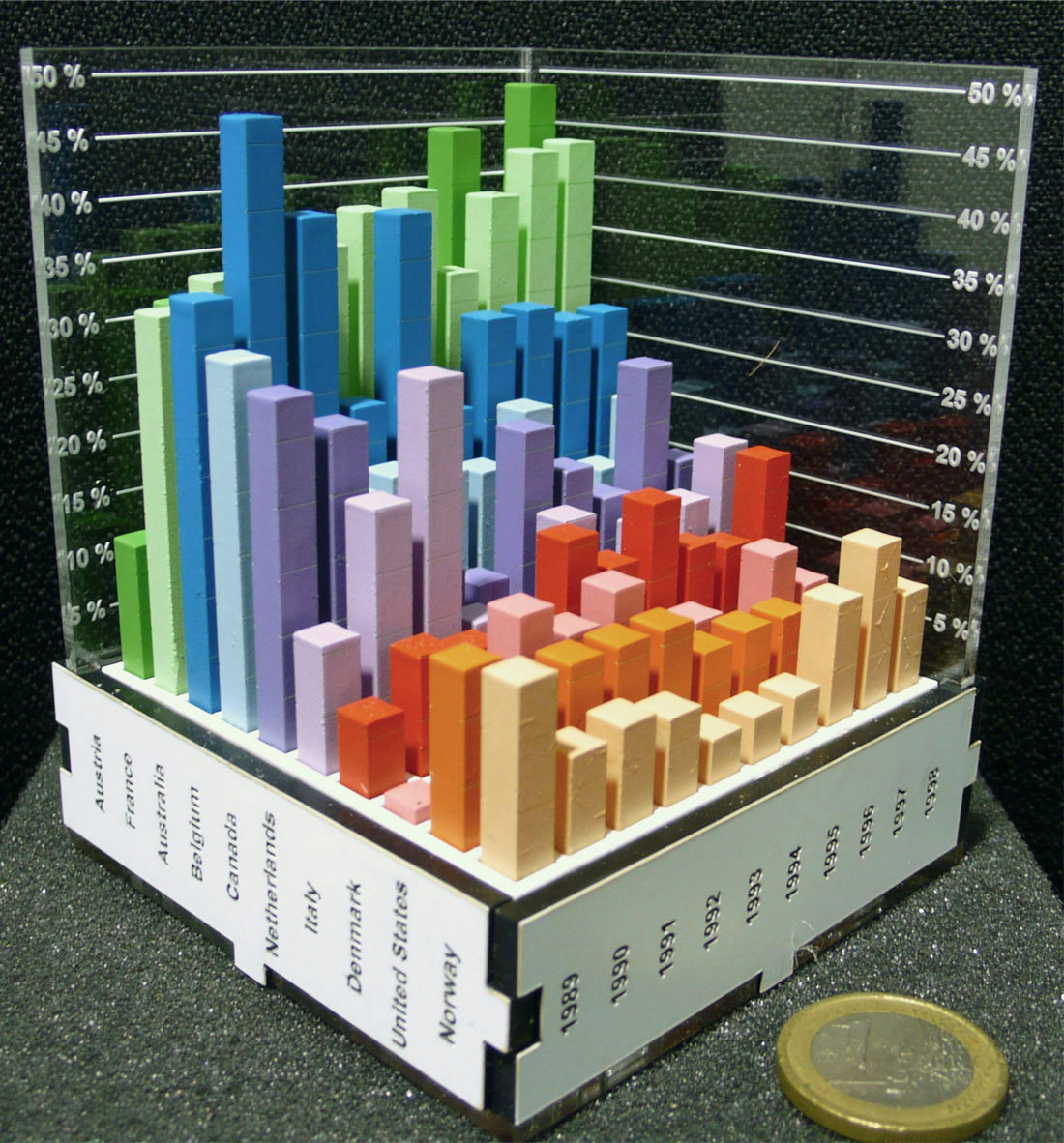}
	\subcaption{}\label{bar}
	\end{subfigure}%
\caption{\subref{map}: Thematic Maps of Germany, Wolf-Dieter Rase; \subref{bar}: Physical bar chart \cite{jansen2013a}}
\end{figure}

In this quadrant, we find physicalization examples that adopt and expand the traditional visual languages of data visualization into physical space. They are based on digital data sets and map their values to material variables. The physicalization community has collected many historical examples of abstract data physicalizations such as spatially stacked charts and histograms, reorder-able matrix displays, spatial networks, and maps (Fig.~\ref{map}).\footnote{See:~\url{http://dataphys.org/list/thematic-maps-of-germany}} The physicalizations often include labeled axes and a color symbology equivalent to 2d charts, intended to make them recognizable as data displays. Jansen and Dragicevic evaluated the efficient comprehension of a fully labeled physical 3d bar chart (Fig.~\ref{bar}) against a 3d rendering and a 2d chart of the same data set, finding that the physicalization was more efficient than the rendering but less efficient than the 2d chart~\cite{jansen2013a}. Interestingly, however, some objects may be recognized as data displays even without labeling --- data-literate observers read the sharply jagged edge of a vertically mounted card as a line-chart with noisy data values along a time axis; spheres connected by arcs or strings as a network diagram. This iconic mode of reception is used in artistic data sculptures and wearable visualizations, which often avoid such signifiers and obscure their data mappings to facilitate a purely formal-aesthetic data experience. The third group in this category includes physical models of spatial phenomena and data sources that are recognizable as such.

\subsection{Epistemological / Relational Examples}

\begin{figure*}[h]
\begin{subfigure}{0.5\textwidth}
	\includegraphics[height=4.7cm]{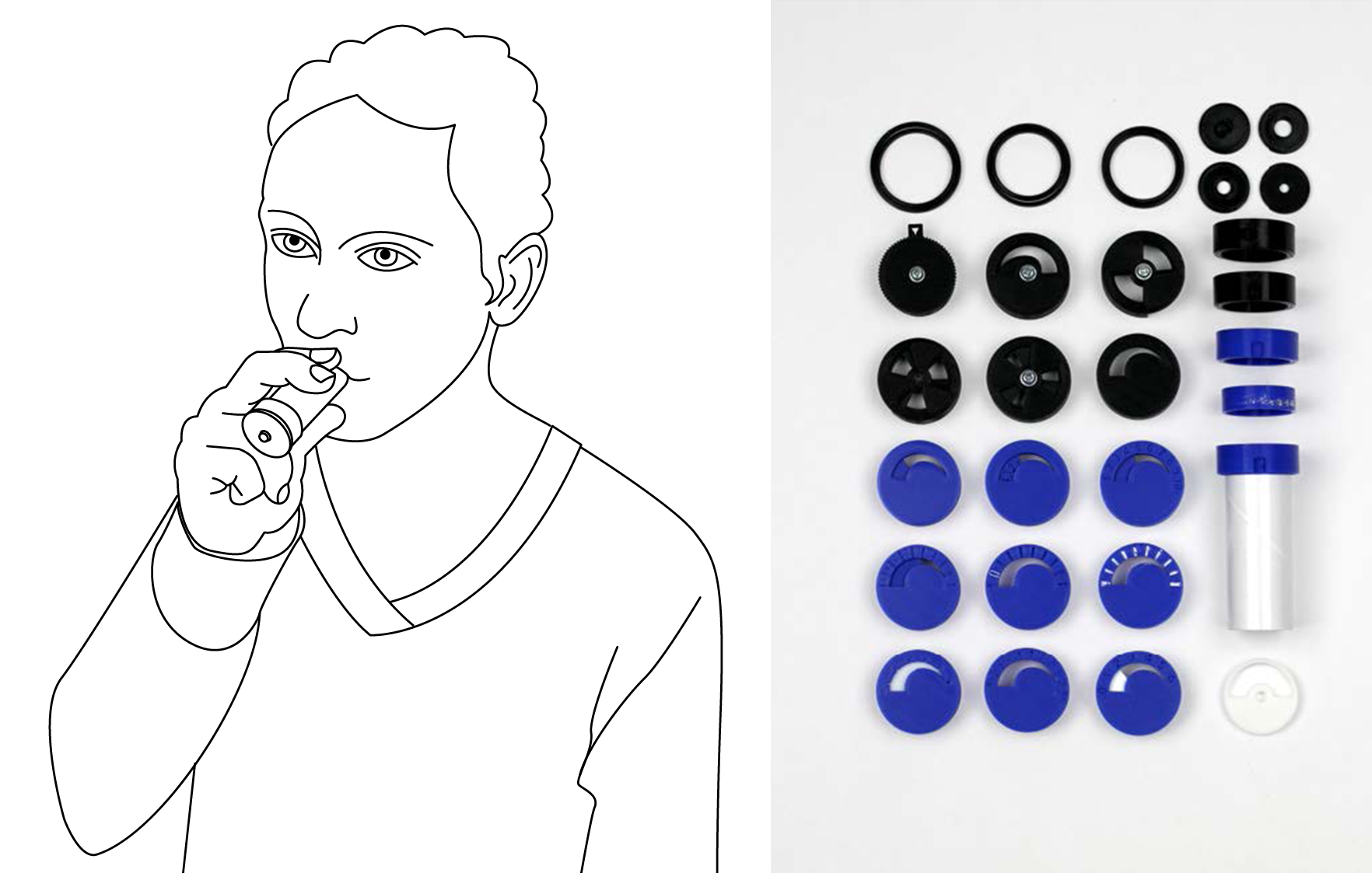}
	\subcaption{}\label{linkner}
\end{subfigure}%
\begin{subfigure}{0.5\textwidth}
    \includegraphics[height=4.7cm]{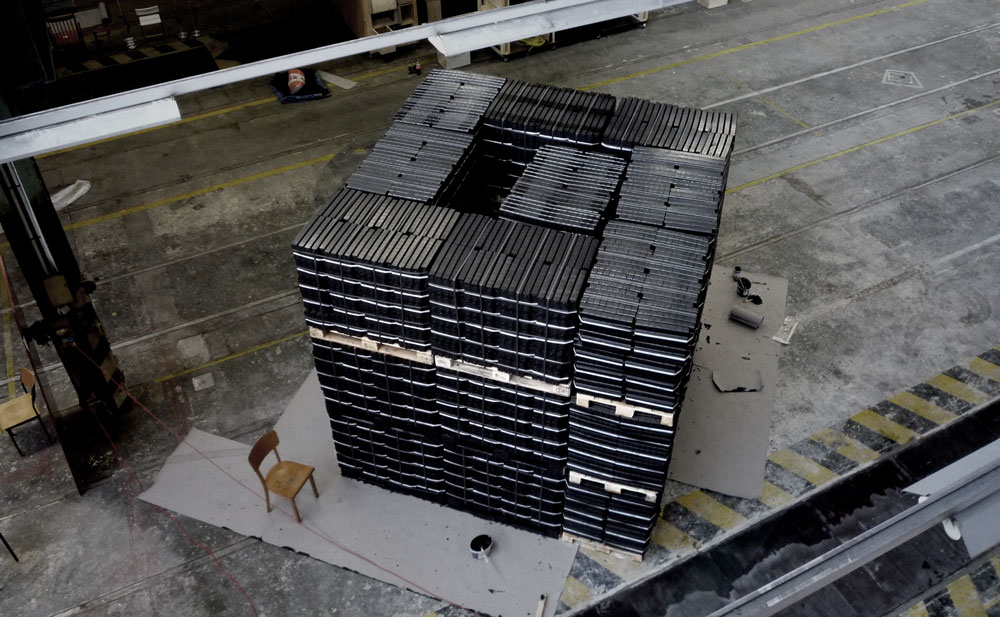}
    \subcaption{}\label{ava}
\end{subfigure}
\caption{Epistemological-Relational examples: \subref{linkner}: Todd Linkner, Breathing Injustice \cite{linkner2020}; \subref{ava}: Michael Saup, The Avatar of the Internet (Source: Michael Saup,~\url{https://1001suns.com})}
\end{figure*}

Examples in the second quadrant are based on symbolic data sets, but express these data by establishing contextual relationships with a specific situation, the recipient, or the process of data collection. They may use data mappings, but don't rely on established conventions as a key to their decoding. Instead, the mappings establish an experiential connection to the phenomenon and the origin of the data set.

How this can take shape can be demonstrated by the recent thesis project "Breathing Injustice" by Todd Linkner (Fig.~\ref{linkner}) \cite{linkner2020}. Interested in the effects of air pollution as an asthma trigger, Linkner physicalized air quality data through a device that the recipient can use to simulate the restrictive effect of pollution on the capacity to breathe of a person with asthma. High concentrations of asthma-related pollution translates to a very constricted tube that makes it more difficult to breathe.

A second example is Michael Saup's sculpture “The Avatar of the Internet.”~(Fig.~\ref{ava} The German artist based his work on calculations of the energy footprint of basic internet services, using the example of the YouTube trailer of the movie “Avatar.” The sculpture consists of a 3x3x3m cube made from lignite coal briquette, “equivalent to the volume of coal that was burnt for the creation of the electrical energy used to serve, transmit and view the online-video-trailer 1 million times,” as the artist explains. Besides its representational function, coal becomes a relational data expression for recipients, who might, as many people in Berlin still do, use coal briquettes to heat their homes and relate to the corresponding amount of energy. The installation could be burned, releasing both energy and pollution that were the subject of the data set.

Other examples include the multi-year \textit{Data Cuisine} project, inviting workshop participants to express a data set through cooking. Many projects go beyond the arbitrary mapping of data to ingredients or creating iconographic representations and instead attempt to establish a relationship with the phenomenon behind the data or its implications. In the example \textit{Eating the Distance}, participants have to drink smoothies through straws that correspond to the distance the fruits had to travel.\footnote{See:~\url{http://data-cuisine.net/data-dishes/eating-the-distance/}}

\subsection{Ontological / Representational Examples}

\begin{figure*}[h]
\begin{subfigure}{0.54\textwidth}
	\includegraphics[height=5.4cm]{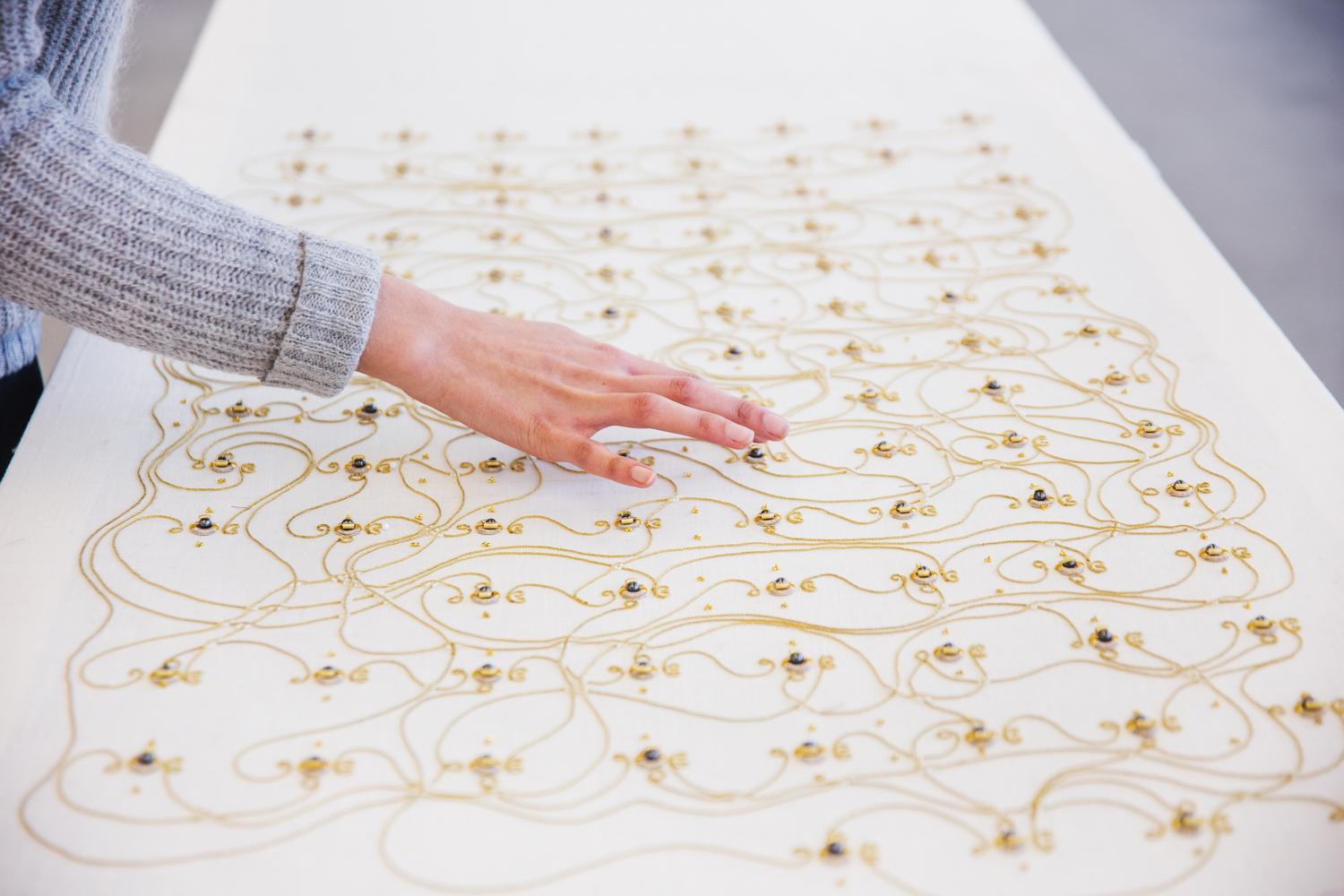}
	\subcaption{}\label{stitching}
\end{subfigure}%
\begin{subfigure}{0.46\textwidth}
    \includegraphics[height=5.4cm]{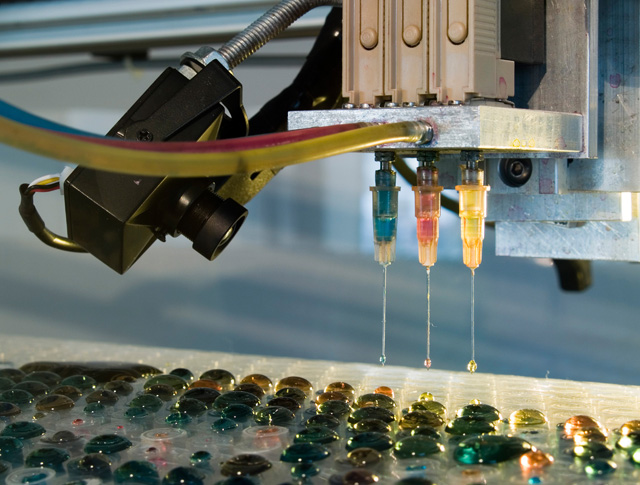}
    \subcaption{}\label{vsstv}
\end{subfigure}
\caption{Ontological-Representational examples: \subref{stitching}: Ebru Kurbak and Irene Posch, The Embroidered Computer \cite{kurbak2018}; \subref{vsstv}: Gebhard Sengmüller, Very Slow Scan Television, 2005 (\url{https://www.gebseng.com})}
\end{figure*}

In the ontological / representational quadrant we find works that derive their informational content from the materials used in the physicalization. They express data, however, by referencing or mimicking established visual languages. This can serve the purpose of providing a familiar point of entry for the recipient or reference the language of another media format rather an external phenomenon. Many media archaeology-themed projects fall into this quadrant, using physicalization methods for a critical analysis of media characteristics.

Ebru Kurbak and Irene Posch's embroidered computer is a functional 8-bit universal electro-mechanical computer, executed as embroideries on fabric using conductive yarns and magnets without using traditional electronic components~(Fig.~\ref{stitching}) \cite{kurbak2018}. The “data” --- the binary states entered by the user --- are processed by the computer’s mechanical-textile gates, which consist of conductive yarn and copper wire wound into coils to form an electromagnet and wrapped around magnetic beads. Its components are presented in a legible arrangement with numerous references to the familiar language of computing. The design allows the audience to recognize the otherwise unusual display as a computer. Ebru Kurbak’s media-archeological explorations of textile computing also involve magnetic yarn as a storage medium for sound, as envisioned a possibility before the invention of the magnetic tape.

A second media artwork that falls into this category is Gebhard Sengmüller's \textit{VSSTV - Very Slow Scan Television} (Fig.~\ref{vsstv}). It is a robotic installation that receives slow-scan television (SSTV) transmissions --- a historic amateur television format --- and recreates the received analog images in an RGB matrix by injecting colored liquid into the air pockets of bubble-wrap packaging film. The result is a barely recognizable image, created by a highly legible apparatus.\footnote{See:~\url{https://www.gebseng.com/02_vsstv}}

\subsection{Ontological / Relational Examples}

\begin{figure*}[h]
    \begin{subfigure}{0.37\textwidth}
    \includegraphics[height=4.25cm]{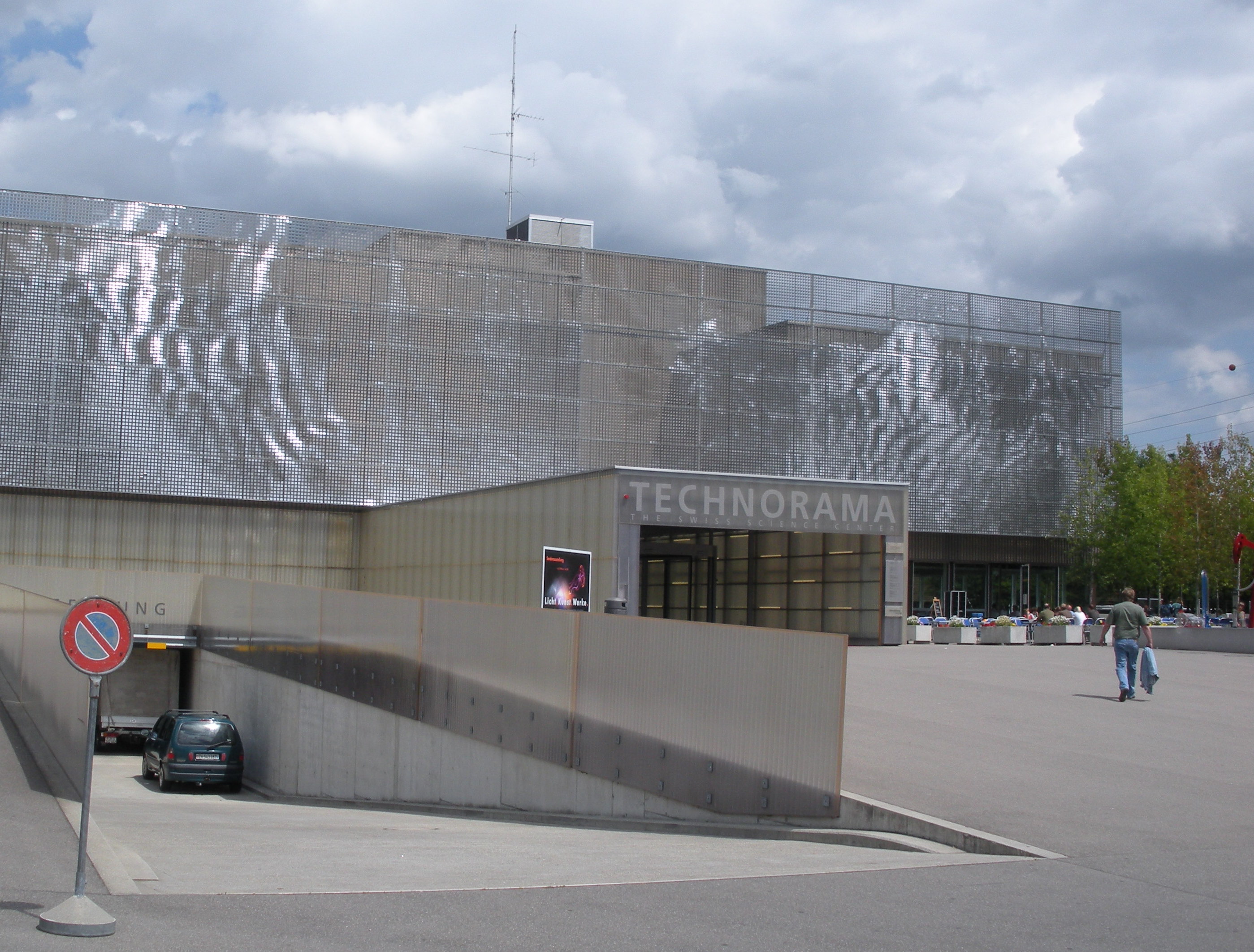}
    \subcaption{}\label{kahn}
    \end{subfigure}%
    \begin{subfigure}{0.19\textwidth}
    \includegraphics[height=4.25cm]{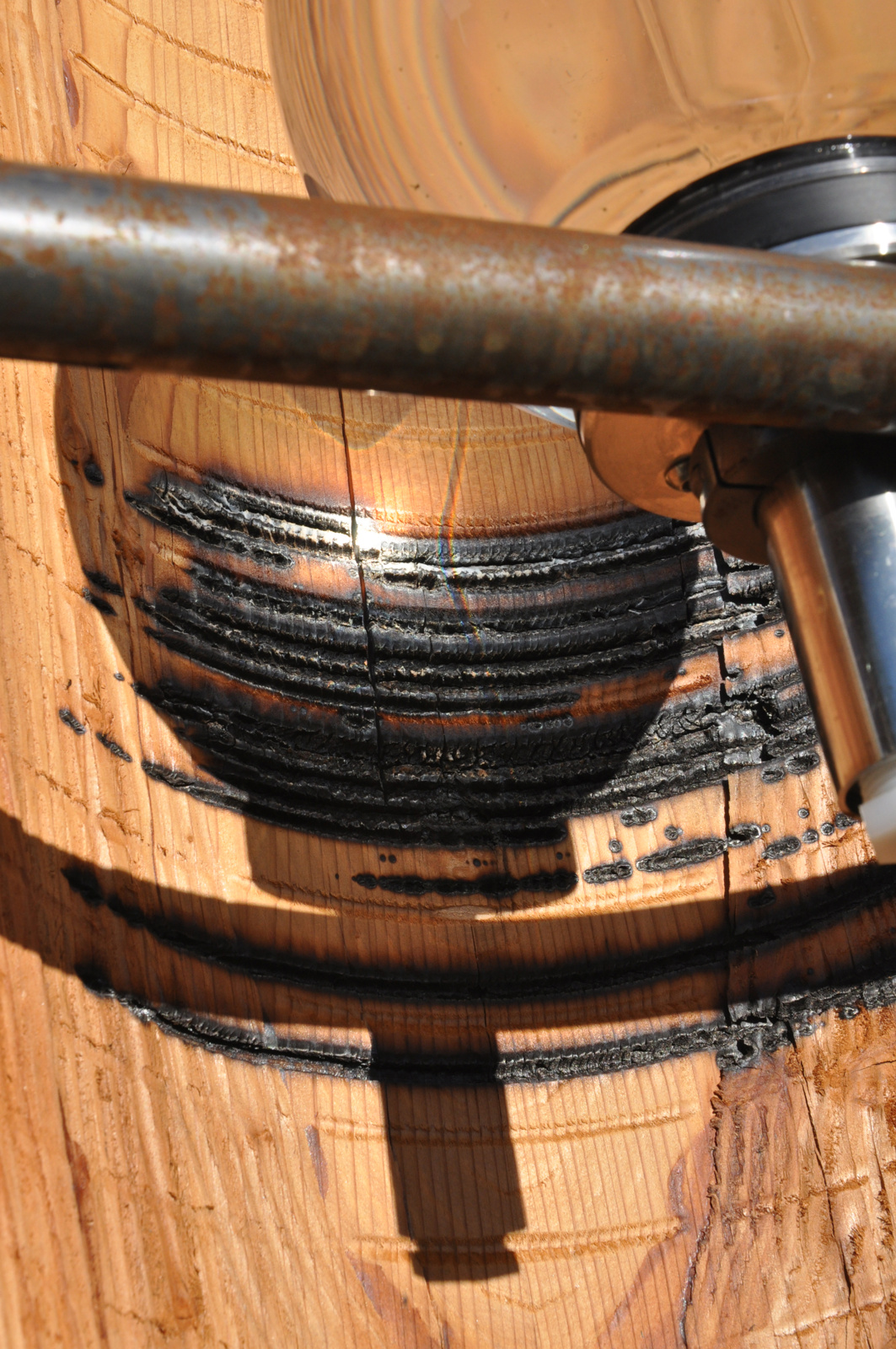}
    \subcaption{}\label{sowers}
    \end{subfigure}%
    \begin{subfigure}{0.4\textwidth}
    \includegraphics[height=4.25cm]{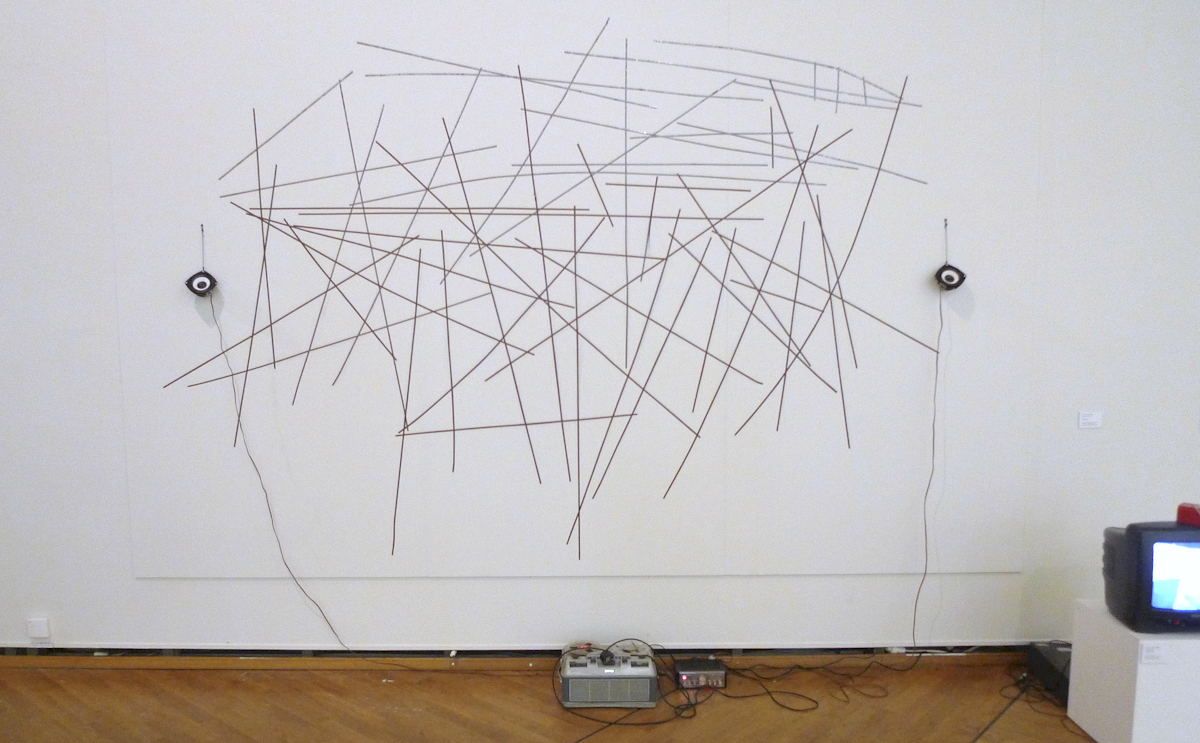}
    \subcaption{}\label{paik}
    \end{subfigure}
    \caption{Ontological-Relational examples: \subref{kahn}: Ned Kahn, Technorama Facade 2002, Winterthur Photo: Javi Rejas, CC BY 2.0 License; \subref{sowers}: Solar Totem (detail), Charles Sowers 2017; \subref{paik}: Random Access, Nam June Paik 1963 Photo: Sascha Pohflepp, CC BY 2.0 License}\label{ont}
\end{figure*}

Examples in this last quadrant operate without traditional data sets and don't use visual languages based on symbolic mappings and representation. They may operate, like \textit{Perpetual Plastic}, by staging physical traces and artifacts as data. They may use traditional forms of media, but deconstruct and defamiliarize them to reveal their inner physical workings. They may also take the form of instruments that reproduce (or reverse-engineer) the data collection process to explain the phenomenon and the material contingencies of data production.

Ned Kahn's wind-activated sculptures visualize the movement of air through an extensive array of small movable aluminum panels that form dynamic patterns in the wind (Fig.~\ref{kahn}).\footnote{See:~\url{http://nedkahn.com}} Charles Sowers' Solar Totems records the daily hours of sunlight in the form of burn marks inscribed into a hollow tree by a spherical lens (Fig.~\ref{sowers}).\footnote{See:~\url{https://www.charlessowers.com}} Both build on classic meteorological instruments such as Alberti's anemometer and Campbell's sunshine recorder. A range of projects aims to make environmental degradation directly perceptible rather than their data representations, including visually marking glacial retreat and sea-level rise or capturing the particles of air pollution. Besides such elemental phenomena, physicalizations in this quadrant can also express social processes, including participatory projects that involve voting with physical tokens that accumulated into visual displays.

Early works in media art have extensively explored the physicality of electronic media and its limits. In \textit{Magnet TV} (1965), Nam June Paik places strong magnets on top of black-and-white TVs, distorting the broadcasted image into abstract shapes that that reveal their technical nature.\footnote{See:~\url{https://whitney.org/collection/works/6139}} In a second piece, Paik dismantles a tape recorder, cutting up the tape and turning it into a wall diagram. The recipient can now manually move the audio head of the tape recorder over the tape strips and listen to their auditory content (Fig.~\ref{paik}). By revealing the materiality of media, Paik's work defamiliarizes and destabilizes representational ideas about data and information.

\section{Discussion and Conclusion}

According to Jansen et al., physical visualizations “map data to physical form”~\cite{jansen2013a}. The data physicalization wiki,\footnote{See:~\url{https://dataphys.org}} however, prominently lists many historical and contemporary projects that are neither based on textual nor digital data. In this paper, I address the underlying desire to take a broader view of data physicalization as a cultural practice by expanding the definition of data. The distinction between epistemological/ontological and representational/relational perspectives on data offers a simple way to think about the role of data in physicalization in terms of data source and data expression.

The simple schema, however, should not be used too literally; it cannot address the complexities of different epistemic worldviews. In general, every visualization or physicalization is an epistemic artifact aimed at building knowledge; ontological perspectives distinguish themselves in nuances by questioning the nature of data rather than interpreting their patterns. Conversely, also born-digital projects can take ontological or relational perspectives; after all, also digital data sets are physical artifacts. The schema is intended as a matrix that allows designers to move beyond dogmatic ideas about data and systematically interrogate which aspects of their physicalizations express and embody data. It may be useful when the abundance of contextual relationships and material variables complicates the interpretation of a physicalization. 

But what if we prefer to avoid these complications and simply apply the notions of data and representation as they are used in visualization? This would mean we have to establish conventions that clearly define how data should be displayed and decoded; in other words, formulate the outlines of a \textit{material literacy}. However, as demonstrated by critiques in related fields such as critical cartography, such conventions fall short of their ambitions of universality and consistency and don't fully account for how most people interpret maps and charts. Just like the standardization in information graphics, such a set of conventions would make us blind to all material properties that are not meant to express data, but establish meaningful contextual relationships. 

Visualization experts rightfully oppose the naive approach to treat color, scale, or length as interchangeable mappings, since each variable has a distinctive relationship with the human body and its sensory capabilities. The evaluation of physicalizations would go beyond questions of ergonomics (e.g. is the user able to differentiate tactile differences?), into the realm of craft (how can the chosen material be manipulated?), cultural references and ethical implications (how does the meaning change when I use wood instead of plastic?). In short, it would be a lost opportunity to look at physicalization just as another application for the traditional visualization toolbox.
\printbibliography
\begin{IEEEbiography}
{Dietmar Offenhuber} is associate professor at Northeastern University and visiting associate professor at Princeton University. He holds a PhD in Urban Planning from the Massachusetts Institute of Technology, a Master of Science in Media Arts and Sciences from the MIT Media Lab, and a Dipl. Ing. in Architecture from the Technical University Vienna. Previously, Dietmar was Key Researcher for visualization at the Ludwig Boltzmann Institute for Media Art Research and professor in Interface Culture at the Art University Linz, Austria.\\
E-mail: d.offenhuber@northeastern.edu
\end{IEEEbiography}
\end{document}